\documentclass[
 aip,
 amsmath,
 amssymb,
reprint,
floatfix,
]{revtex4-1}

\usepackage[utf8]{inputenc}
\usepackage{lineno,hyperref}
\usepackage{graphicx}% Include figure files
\usepackage{dcolumn}% Align table columns on decimal point
\usepackage{bm}% bold math
\newif\iffinal
\iffinal
  \newcommand\ian[1]{}
\else
   \newcommand\ian[1]{{\color{blue}[Ian: #1]}}
\fi

\usepackage{algpseudocode,algorithm,algorithmicx}
\usepackage{hyperref}

\usepackage[utf8]{inputenc}
\usepackage[T1]{fontenc}
\usepackage{mathptmx}
\usepackage{xcolor}
\usepackage{todonotes}

\begin{document}

\title[]{What is missing in autonomous discovery: Open challenges for the community}
%  alphabetical.https://www.overleaf.com/project/630e7f3f6803066427437d86
\author{Phillip~M.~Maffettone}
\email[]{pmaffetto@bnl.gov}
\affiliation{National Synchrotron Light Source II, Brookhaven National Laboratory, Upton, NY 11973, USA}
\affiliation{BigHat Biosciences, San Mateo, CA 94403, USA}
% ORCID: 0000-0001-7173-7972

\author{Pascal~Friederich}
\email[]{pascal.friederich@kit.edu}
\affiliation{Karlsruhe Institute of Technology, Karlsruhe, Germany}
% ORCID: 0000-0003-4465-1465

\author{Sterling G. Baird}
\affiliation{University of Utah, Salt Lake City, UT 84108, USA}
% ORCID: 0000-0002-4491-6876

\author{Ben Blaiszik}
\affiliation{University of Chicago, Chicago, IL 60637, USA}
\affiliation{Argonne National Laboratory, Lemont, IL 60439, USA}
% ORCID: 0000-0002-5326-4902

\author{Keith~A.~Brown}
%\email[]{brownka@bu.edu}
\affiliation{Boston University, Boston, MA 02215, USA}
% ORCID: 0000-0002-2379-2018

\author{Stuart~I.~Campbell}
%\email[]{scampbell@bnl.gov}
\affiliation{National Synchrotron Light Source II, Brookhaven National Laboratory, Upton, NY 11973, USA}
% ORCID: xxxx-xxxx-xxxx-xxxx

\author{Orion~A.~Cohen}
%\email[]{orion@lbl.gov}
\affiliation{Materials Science Division, Lawrence Berkeley National Laboratory}
% ORCID: 0000-0003-3940-2456

\author{Tantum Collins}
\affiliation{Centre for the Governance of AI, Oxford, UK}

\author{Rebecca~L.~Davis}
%\email[]{rebecca.davis@umanitoba.ca}
\affiliation{University of Manitoba, Winnipeg, Canada}
% ORCID: xxxx-xxxx-xxxx-xxxx

\author{Ian~T.~Foster}
%\email[]{foster@anl.gov}
\affiliation{Argonne National Laboratory, Lemont, IL 60439, USA}
\affiliation{University of Chicago, Chicago, IL 60637, USA}
% ORCID: 0000-0003-2129-5269

\author{Navid~Haghmoradi}
%\email[]{navid.haghmoradi@kit.edu}
\affiliation{Karlsruhe Institute of Technology, Institute of Nanotechnology, Hermann-von-Helmholtz-Platz 1, 76344 Eggenstein-Leopoldshafen, Germany}
% ORCID: xxxx-xxxx-xxxx-xxxx

\author{Mark~Hereld}
%\email[]{hereld@anl.gov}
\affiliation{Argonne National Laboratory, Lemont, IL 60439, USA}
\affiliation{University of Chicago, Chicago, IL 60637, USA}
% ORCID: 0000-0002-0268-2880

% \author{Howie~Joress}
% %\email[]{howie.joress@nist.gov}
% \affiliation{Materials Measurement Science Divison, National Institute of Standards and Technology, Gaithersburg, MD 20899, USA}
% ORCID: xxxx-xxxx-xxxx-xxxx

\author{Nicole~Jung}
%\email[]{nicole.jung@kit.edu}
\affiliation{Karlsruhe Institute of Technology, Institute of Biological and Chemical Systems, Hermann-von-Helmholtz-Platz 1, 76344 Eggenstein-Leopoldshafen, Germany}
% ORCID: 0000-0001-9513-2468

\author{Ha-Kyung~Kwon}
%\email[]{ha-kyung.kwon@tri.global}
\affiliation{Toyota Research Institute, Los Altos, CA 94022, USA}
% ORCID: 0000-0002-9351-4806

\author{Gabriella Pizzuto}
%\email[]{gabriella.pizzuto@liverpool.ac.uk}
\affiliation{University of Liverpool, Liverpool, L69 3BX, UK}
% ORCID: xxxx-xxxx-xxxx-xxxx

\author{Jacob Rintamaki} 
%\email[]{jrin@stanford.edu}
\affiliation{Stanford University, Stanford, CA 94305, USA} 
% ORCID: xxxx-xxxx-xxxx-xxxx 

\author{Casper~Steinmann}
%\email[]{css@bio.aau.dk}
\affiliation{Department of Chemistry and Bioscience, Aalborg University, 9220 Aalborg, Denmark}
% ORCID: 0000-0002-5638-1346

\author{Luca~Torresi}
%\email[]{luca.torresi@kit.edu}
\affiliation{Karlsruhe Institute of Technology, Institute of Theoretical Informatics, Engler-Bunte-Ring 8, 76131 Karlsruhe, Germany}
% ORCID: xxxx-xxxx-xxxx-xxxx

\author{Shijing~Sun}
%\email[]{shijing.sun@tri.global}
\affiliation{Toyota Research Institute, Los Altos, CA 94022, USA}
% ORCID: xxxx-xxxx-xxxx-xxxx       

\begin{abstract}
Self-driving labs (SDLs) leverage  combinations of  artificial intelligence, automation, and advanced computing to accelerate scientific discovery. 
The promise of this field has given rise to a rich community of passionate scientists, engineers, and social scientists, as evidenced by the development of the Acceleration Consortium and recent Accelerate Conference.
Despite its strengths, this rapidly developing field presents numerous opportunities for growth, challenges to overcome, and potential risks of which to remain aware. 
This community perspective builds on a discourse instantiated during the first Accelerate Conference, and looks to the future of self-driving labs with a tempered optimism.
Incorporating input from academia, government, and industry, we briefly describe the current status of self-driving labs, then turn our attention to barriers, opportunities, and a vision for what is possible.  
Our field is delivering solutions in technology and infrastructure, artificial intelligence and knowledge generation, and education and workforce development. 
In the spirit of community, we intend for this work to foster discussion and drive best practices as our field grows.

\end{abstract}
\maketitle

%\linenumbers

\section{Introduction}
%%% Outline %%% 
% 1. Cambrian explosion of SDLs
% - Promises from reviews and bold claims( discovery, to tech readiness, to industrial capacity)
% - Elicitinin a suite of funding and dedicated platforms (Journals/confs) (Acc, DOE, DOD, DARPA, Canada, Journals)
% - not quite a scientific revolution, but an industrial one that will change the pace of scientific ones

% 2. What is an SDL
% - Stick a graphic in here
% - Disambiguation of language
% - Automation; Autonomous experimentation; SDL; MAP; 

% 3. Big chonking thesis paragraph
% - What this perspective is, and what it is not. 
% - We gathered
% - We bantered
% - We hit on these sections
% - We shared with community
% - Impassioned statement about the future 

Scientific experimentation and discovery is teetering on the precipice of a new industrial revolution. 
Acceleration of science by combining automation and artificial intelligence (AI) has begun to revolutionize the structure of scientific experiments across physics \cite{Roccapriore_2022}, chemistry \cite{abolhasani2023rise,shields2021bayesian,hase2019next}, materials science \cite{tabor2018accelerating}, and biology \cite{Narayanan_2021}. % cite unique reviews here for each
The integration of high-throughput experimentation, AI, data science, and multi-scale modeling have spawned great interest \cite{Seifrid_2022}, notable results \cite{Stach_2021}, and substantive expectations.
These expectations include acceleration of experimental throughput, new discoveries, technological readiness, and industrial adoption.
Such excitement has elicited a suite of conferences (from a 2018 North American workshop culminating in the first Mission Innovation report \cite{missioninnovation2018} coining the name Materials Acceleration Platform (MAP), to the most recent Accelerate Conference), dedicated publication platforms, and increased funding from governments and the private sector. 
Furthermore, as a link between algorithms and the real world, self-driving labs (SDLs) are a prerequisite for further advancements of autonomous and AI-driven research, as targeted by, for example, the Turing AI Scientist Grand Challenge as a forward-looking roadmapping effort\cite{kitano2021nobel}.
While this rapid community advancement may not yet constitute a scientific revolution \cite{shapere1964structure}, it does initiate a technical revolution that will likely change the pace at which we see scientific breakthroughs. 

\par

A self-driving laboratory (SDL) can be described as a scientific system that performs autonomous experimentation (AE).
That is, it uses automation and AI to operate and select each successive experiment, without requiring human intervention.
Several other terms  are commonly used in this domain which are worth disambiguation from SDLs.
High-throughput experimentation (HTE) applies automation technology or engineering practices to increase data generation rates, but these experiments are often fully designed by human experts or predefined at the start of an experimental campaign \cite{Stach_2021}. 
A common term used for SDLs in the materials science domain is materials acceleration platform (MAP) \cite{missioninnovation2018}. 
The major differentiator between platforms is the degree of autonomy imbued in the system---and increasingly, the degree of human-AI collaboration. 
In the following, we will use the term SDL and focus broadly on (semi-) automated platforms that accomplish (high-throughput) experiments, process and analyse results autonomously, and use that analysis to guide future experiments. 
While an SDL can be typified by a closed loop of synthesis and analysis unit operations \cite{Burger2020}, a single SDL can be incorporated as a unit operation inside a larger SDL, so long as it meets the criteria above \cite{MaffettoneMASS}. 
\par
% THIS IS A GOOD PLACE FOR A SCHEMATIC FIGURE :)
% \missingfigure{THIS IS A GOOD PLACE FOR A SCHEMATIC FIGURE}
\begin{figure}
    \centering
    \includegraphics[width=\linewidth]{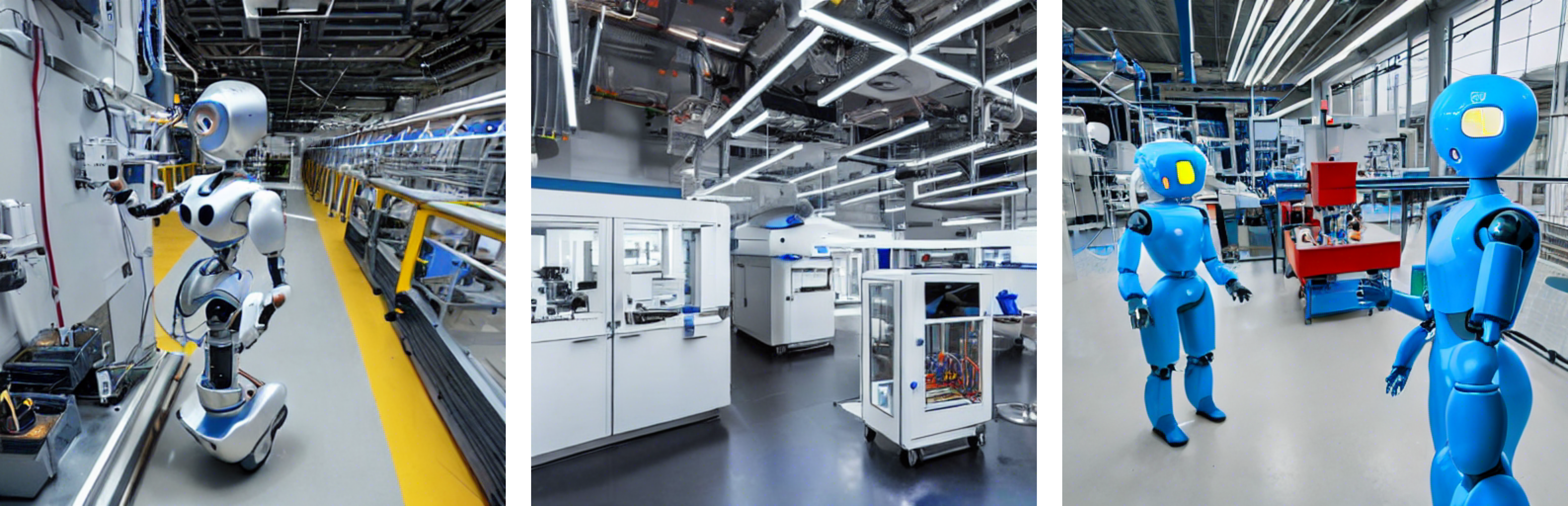}
    \caption{A triptych of stable-diffusion generated images describing a self-driving lab for autonomous scientific discovery.}
    \label{fig:abstract}
\end{figure}

Herein, we describe a community perspective on the state of SDLs, focused on open challenges and concerns for the future.
This perspective incorporates insights from academia, government, and industry, including both users and developers of SDLs.  
This work grew organically from discussions started at the first Accelerate Conference in September 2022, and is intended to foster an ongoing discourse to influence the field of autonomous experimentation. 
%%%% For arXiv draft %%%%
As such, we have posted a first draft of this work as a pre-print to encourage new feedback, insights, and collaboration that will shape an updated version of the perspective.
We invite interested readers to reach out to the authors to engage in this process and to join the community discourse. 
%%%% For arXiv draft %%%%
While we will first describe the current state of the field, this is by no means a comprehensive review, and we encourage the reader toward general reviews of SDLs and autonomous experimentation \cite{Stach_2021}, as well as summary articles in scientific sub-disciplines \cite{Narayanan_2021,Roccapriore_2022,abolhasani2023rise,shields2021bayesian,hase2019next,tabor2018accelerating}.
Following this, we will turn our attention to barriers and opportunities associated with data, hardware, knowledge generation, scaling, education, and ethics.
As the field of autonomous experimentation grows and SDLs become more common, we hope to see rapid growth in scientific discovery. 
In the community approach to vision that follows, we look to ethically and equitably accelerate this growth and adoption.

%%% Local Variables: 
%%% mode: latex
%%% TeX-master: "./main.tex"
%%% End:
\section{Data, data, everywhere, and not a piece to parse}\label{sec:data_sharing}
% Drawing text from data_sharing_incentives.tex

% 1. Best practices
% - what is FAIR, why it is good
% - what kind of infrastructure is used
% - where are attempts to emulate
% 2. Bariers in data sharing
% - pop something here on infra
% 3. Incentive ideas
% - elude to education and access on the infra angle 

A core difference between research using SDLs and conventional research is the amount and structure of data that are generated.
Compared to human-guided experimentation, SDLs can enable rapid experiments, with improved reproducibility, and with automatically generated metadata.
This presents many opportunities to reduce the barrier and overheads to share, collect and use data globally. 
At the same time, it leads to challenges in terms of storage, analysis, and even interpretation of data.
Perhaps one of the unique opportunities afforded by SDLs is that data could be shared in real time with the community while preserving its provenance.
In the following, we will discuss best practices for data sharing, but also existing barriers and incentives to lower and overcome those barriers. 
Data is foundational to SDLs and leveraging the full potential of data-driven approaches in science remains an area of rich opportunity.

% \subsection{Best practices and barriers in sharing data}
% some definitions and objectives
Best practices of data sharing and publication are summarised in the FAIR principles~\cite{wilkinson16, brinson2022fair}, i.e., data that are  findable, accessible, interoperable, and reusable.
\textit{Findable} means that data should be easy to find and access, with clear and consistent metadata, and data identifiers.
\textit{Accessible} means that data should be openly available through a trustworthy and stable digital repository with a clear and simple access mechanism.
\textit{Interoperable} means that data should be structured with common standards, shared vocabularies, and standardized formats that enable integration and reuse.
\textit{Reusable} means that data should be well-described with provenance and licenses, allowing it to be used and cited in new research and for new purposes.
FAIR methods are most effective when applied continuously during research, rather than only at the moment of data publication~\cite{dempsey2022sharing}.
These principles enable the linking of many experiments and simulations, and enable SDLs to save resources, leverage existing knowledge, and utilize synergies.
% transition to barriers and challenges
Despite the successful implementation of FAIR principles by some exemplary practitioners \cite{escalate, Blaiszik_2019}, there remain numerous technological and cultural barriers to FAIR data sharing.
It is thus important to be forthcoming about these barriers, as we look to understand what strategies or incentives are viable for improving the overall data publishing landscape.
\par

%\subsection{Current challenges and barriers in data sharing}
% technical challenges
% data acquisition and digitization
Primary technical barriers to entry center around data acquisition and digitization.
An early challenge is the movement of data from unit operations to data storage infrastructure, databases, or long-term repositories \cite{bluesky, opc, sila, patterns}.
While electronic lab notebooks can help integrate persistent manual processes, data acquisition in SDLs is mostly challenged by lack of accessibility, as many scientific instruments still produce data in obfuscated proprietary or binary formats \cite{sila} and data often remains undiscoverable at the point of data generation. 
Developing open interfaces to hardware is an ongoing and increasingly solved task, which will enable open software to collect and manage the data as it is produced.
Data, i.e. not only characterization data but also data regarding processes, should ideally be recorded as close to the source as possible. If manually recorded later, this will not only increase the effort of data collection but also potentially lead to data loss due to incompleteness and missing standardization. Again, electronic lab notebooks connected to electronic equipment are one way to digitize (process) data directly during generation, while removing the effort of additional book-keeping.
\par

% RDM 
Once created, digital data can easily be stored on a hard disk as a series of named files, yet this approach does little to ensure that the data can be found, understood, and reused in the future by collaborators, the community, or machines, without insight from the original data creators.
This problem is further exacerbated by the fact that most researchers are accustomed to the file systems of their laptops, but lack the background, training, or incentive to use shared community databases, common metadata standards, unique data identifiers, and well-defined vocabularies \cite{Pelkie_2023}.
Furthermore, there are additional technical challenges connected to choosing appropriate databases (hierarchical, relational, non-relational) and data storage with appropriate data protection, safety, and maintenance (local or cloud).  
The development of research data management (RDM) tools, often driven by bottom-up initiatives within domain specific communities, help overcome these hurdles.
RDM developments can facilitate FAIR data by providing appropriate shared metadata standards, domain-specific vocabularies and ontologies, and also software and storage solutions.
An example RDM system that provides some of this functionality is contained in the \textit{Bluesky} project,  which implements a standard ``Data API'' that is exposed to users and other systems rather than requiring knowledge of internal implementation details on where and how the data is stored~\cite{bluesky}.
\par

% interoperability and data formats
The last set of technical barriers relates to data interoperability, which also extends to the homogenization of heterogeneous data sources.
Relating data from disparate equipment, including distributed instrumentation, multi-fidelity probes, or even simulations, requires some knowledge of the core metadata used to describe each relevant experiment.
Unfortunately, constructing these relationships requires some degree of standardization and is highly discipline dependent.
Crystallography provides an excellent example of standardizing data formats for interoperability (protein data bank and crystallographic information files) \cite{pdb, cif}.
While this standardization provides a means of comparing measurements, it does not provide rich metadata for samples.
Data validation and schema \cite{Breck_2019} can be used to create interoperable sample data.
Overcoming this challenge will take effort, open communication, and many revisions.
Even in a single lab, lean engineering practices should be applied to a regularly updated and versioned data model (e.g., ambient humidity or materials batch numbers may not be required data fields until a keen researcher notes them as important exogenous variables).
We therefore encourage the community to develop, collaborate on, and publish data models, particularly in a version-controlled manner.
Even highly specific data models can prove impactful and be improved upon.

% databases
An open challenge in many domains---or sub-domains---remains the creation and support of easy-to-use, open-access domain-specific repositories and databases.
Similar to the domain-specific nature of sample data models, the possibilities are innumerable and can start small and iterate. 
It is important to note the difference between generic databases and repositories such as FigShare and GitHub, currently only domain-specific databases provide data with enough structure to be reused within the community \cite{tremouilhac2020repository}.
To improve discoverability, there are a number of FAIR data databases and repositories in the materials science community that may prove useful to collecting SDL data and for the SDL community to build upon, including the Materials Project~\cite{materialsproject}, AFLOW~\cite{aflow}, the Materials Data Facility~\cite{mdf1, mdf2}, OQMD~\cite{oqmd}, and NOMAD~\cite{nomad}. However, none have yet been closely integrated with SDL.
As more databases become available, a new opportunity will be present in building infrastructure to interoperate between them or merge them under single data models. 
This again relates back to challenges with the ingestion of standardized data formats for various hardware vendors to seamlessly relate measurements \cite{herring2020beep}.
Quality control and continuous integration are challenging tasks even for data workflows from a single source \cite{BigHatCICD}; as interoperation grows, so will the challenges for quality control and integration.
This will be a welcome opportunity and a hallmark of community progress.
It can be addressed by first building internal trust in data through QC pipelines that are incorporated in a data model.
Cloud tools such as AWS and Azure, as well as workflow managers \cite{chard2023globus} can be used to build automation into the construction, maintenance, and QC of these databases. 

Aside from the substantial technical challenges of building a robust data infrastructure, there are many cultural hurdles.
Some are centered around sharing proprietary data that represents a material value to its creators. 
Others reflect researchers' fear of being "scooped" or yielding a competitive advantage.
While technical issues can be solved with clever engineering and sufficient funding, cultural issues cannot. 
Instead, workshops, conferences, training programs, and higher education serve as the engines of cultural change.
We suggest conceptualizing improved data management as a socio-technical transition pushing against both technical and cultural lock-in to existing practices.

%\todo{Finish, explain and rationalize}

\subsection{Suggestions for data sharing incentives}

We have identified three strategies for incentivizing the production of FAIR data, putting into focus the human researcher, rather than MAP technology:
\begin{itemize}
\item Reducing friction
\item Providing rewards
\item Demanding requirements
\end{itemize}
\par

Creating the technical means to share data in a simple way is a necessary condition to allow researchers to share data in a sustainable way. Reducing friction means making it simple and effortless for researchers to upload their data to the appropriate places.
Whether this is the time it takes to create a DOI for data, or the effort needed to validate the data into a given database schema, people will be more likely to engage with a tool if it does not feel like work.
We encourage the ``customer development'' required for creating an appropriate user interface for data tools \cite{Maffettone_2022_neurips}.
This ties into the above discussion of creating and maintaining field-specific databases, which are easily accessible and also integratable into SDLs to further enable the automation of data publishing pipelines.
If these are accessible with a continually refined user interface, the community will be more likely to contribute their data.

Even if they are easy to use, technical tools are not enough to create a sustainable development toward openly sharing data. 
To encourage researchers to share data more broadly, we believe that data sharing should be met with recognition and rewards.
The recognition could take the form of data citations, which are tracked by the statistics of access and downloads.
Such recognition would directly create incentives to publish in databases that impose high standards to ensure high quality, reproducibility, and documentation of data, as such databases will be used and thus cited more by the community.
Furthermore, this can incentivize the publication of ``negative-data'' \cite{Raccuglia_2016}. Broad data availability can further lead to citations by researchers without access to labs and HPC infrastructure.
Citable data can then lead to the definition of new metrics.
Consider, for instance, the social and professional impact of the h-index. 
While there are systematic challenges associated with over-reliance on a single metric, the h-index provides a more holistic---if flawed---means for measuring impact.
We envision a complementary metric for data (e.g., a d-index), that could be built off of the unique identifiers for data in a database, allowing data to be referenced digitally in publications.
While new metrics bring new concerns, integrating these metrics into the traditional advancement criteria at research institutions would produce a dramatic cultural shift.
\par

An alternative approach to creating incentives in the currently existing research reward system centered around publications and citations is the use of automated papers that are published regularly, listing all recent contributors to a public database. 
This can even be transferred to track software impact, e.g., publishing a citable list of contributions to given repositories.
This idea is not entirely new, as large software tools regularly publish reports of new  versions with all contributors as authors \cite{numpy}.
However, the main outcome of this strategy is converting data contributions into publications/citations, and pivots the credit mechanism to an already imperfect set of metrics. 
While the career advancement for a scientific software engineer can be as closely tied to their publication record as their git history, no such paradigm has been publicly recognized by an R1 university, especially not in the context of data provenance.
Thus, we think creative solutions to provide micro-incentives to sharing data and exercising good data practices will be a critical need in making progress in this field and encourage social scientists to consider this research problem.
\par

In continued circumstances where a reward infrastructure proves lacking, we encourage mandates through peer review and funding agencies.
This policy can be enacted at the journal or editor level, or be enforced at the level of individual peer review, as FAIR data are a reasonable pre-requisite for reproducibility.
Some journals already enforce code review as part of their process for scientific software, and it is sensible to enforce data review for papers that describe large datasets.
With regards to funding, many government funding agencies require a data plan, albeit, enforcement is critical.
Furthermore, data plans often only include positive results.
Public access to data created using public funds is paramount and should be rigidly enforced by funding bodies. 
Data plans, however well-designed, are not useful if they are not used. 
There is also no reason why this must be done from scratch: for example, a data management framework could be developed and highlighted alongside a new dataset.
Best practices will naturally differ between communities with different measurement techniques, instrumentation, and figures of merit.
It is therefore advisable that data management frameworks be designed through active collaboration between scientific communities and funding agencies.

%%% Local Variables: 
%%% mode: latex
%%% TeX-master: "./main.tex"
%%% End:
\section{Integrating hardware into SDLs}\label{sec:hardware}
The experimental apparatus that constitutes the physical embodiment of the SDL provides its own set of challenges.  
While data and software solutions are reusable across a large swath of the research landscape,  hardware advancements must be capable of handling the specifics of the research problem at hand. 
Further, physical platforms are generally the most capital-intensive part of an SDL, including  the costs of the  scientific instrumentation required for a given experimental workflow, and often custom automation hardware.
Therefore, we stress an objective that the hardware powering SDLs generalise across different experiments to make the investment reasonable. 

% \subsection{State-of-the-art technology and challenges}
There are three common approaches to hardware in current SDLs: building hardware from individual components (e.g., motors, pumps, controllers, detectors); using workcells (i.e. integrated systems which bring together automation equipment,  analysis tools, and software, to accomplish rigid predefined tasks); and integrating unit operations with an anthropomorphic robotic platform.
The from-scratch approach is most common in specialist equipment, such as beamlines \cite{Maffettone_2022_neurips, bluesky}, new microscopes \cite{Hohlbein_2022}, or specialist synthesis approaches \cite{Tabor_2018}; however, growing maker communities have driven some build-your-own workcells similar to 3D printer technology \cite{jubilee}.
Workcells first came to use in high throughput biological applications \cite{King_2009}, and now have commercial providers across the physical sciences. 
These are relatively rigid unit operations for a given experiment type, although it has been demonstrated that a robotic arm can be integrated internally which could provide more flexibility \cite{macleod2020self}.
To add even more flexibility, the use of mobile robots for transporting sample vials and using equipment across different laboratory stations has been demonstrated \cite{Burger2020}, in addition to the usage of heterogeneous robotic platforms depending on the laboratory tasks~\cite{Fakhruldeen2022}.
As such the current state of the art for laboratory automation varies depending on the commonality of the task to automate, and the commercial demand for products; a generic liquid handler has a much broader market appeal than a domain-specific workflow. 
\par

Although building hardware is hard, it pales in comparison to the challenge of integrating hardware.
This has amassed public \cite{sila, quigley2009ros, bluesky, escalate, Fakhruldeen2022} and private efforts.
Developing a hardware approach for an SDL involves a combination of deciding on the best tools for the laboratory task and building common interfaces for those tools.
As an example, we contrast a synthesis workflow that can be accomplished in a commercially available workcell or using custom equipment, an analysis workflow that depends on advanced detectors, and a sample management workflow that uses robotic arms. 
The synthesis workcell may include a software interface that cannot be rebuilt, whereas the custom synthesis equipment will need to choose an effective software interface. 
Advanced detectors on the other hand are commonly integrated using open software tools \cite{epics, tango}.
Lastly, robotic arms are traditionally driven using middleware that, through already available libraries, have in-built motion planning, low-level controllers, and perception \cite{quigley2009ros}.
While these are fundamental for autonomous robotic platforms, given that most synthesis or analysis workflows are carried out in open-loop, these functionalities have not been fully exploited. 
This then raises the question of how to best integrate these disparate systems without attempting the mammoth task of rebuilding them all using a single software tool.
\par

We encourage the development and reuse of open-source, non-proprietary hardware communication, as well as interfaces between those common platforms.
This would facilitate sample exchange across multiple commercial experimental tools, in conjunction with bespoke tooling, as well as improving knowledge transfer, reproducibility, and generalization to new SLDs. 
Interfaces should make use of industry-standard message bus technology, that will enable both local and cloud operation.
Moreover, these interfaces are compatible with streaming data, that enable \emph{in situ} and real-time measurements with automated data-processing pipelines.
The materials community has called for investment in ``the redesign of microscopes, synchrotron beamlines, and other sophisticated instrumentation to be compatible with robotic sample handling---akin to the multi-plate-handling robots in the bio-community.''\cite{Stach_2021} Such innovation would be empowered by the adoption of open frameworks and message buses (e.g., by integrating a ROS-enabled robot, a Bluesky-enabled beamline, and SILA-enabled unit operations).
\par

The economy of public research requires that SDL platforms---or at least their components---be reusable beyond the scope of a single research project.  
One approach to this is to develop modular systems that can be added to over time to meet the new demands of new research questions.
This would be supported by  a common physical sample interchange environment.
The details of sample interchange greatly depends on the form factor of the material being measured.  
\par

Samples can generally be clustered into three varieties: liquid, bulk solids (including thin films on substrates) and powders.  
Liquid handling is the most common in current SDL setups, having been pioneered by the bio-pharmaceutical sector.  
Samples can be physically handed between instruments in vessels, moved through the use of pipetting, or directly pumped through piping. 
Handling of bulk solids is also relatively straightforward.  
Robotic arms are able to move samples between processing and characterization tools.  
The sample surface is also readily available in this form factor for characterization.  Each sample may be homogeneous or contain multiple sub-samples.
Powders are perhaps the most challenging to deal with using automation.  
While precursor powders can be readily dispensed, powders as a product can be difficult to handle.  
While powders can be moved in a vial, this is often not a form factor that is amenable for characterization.  
For example, X-ray diffraction typically requires creating a flat surface or packing into a capillary. 
Conversely, powders for catalysis or absorption of gasses typically require these powders to be packed into columns for testing.  
Further complicating this process is the fact that powders can have a great variety of flow properties, requiring adaptive manipulation.
We encourage robotics and mechatronics research in this area, as automation of powder handling would solve a particularly impactful tactile challenge.

% It is worth dedicating special attention to the area of applied robotics in SDLs. \todo{Gabriella, can you take a look here?}
\subsection{Applying Robotics to SDLs}
Deploying anthropomorphic robotic systems in SDLs is a promising area of research in that it enables the use of existing or standard instrumentation, empowers human collaboration for non-automated tasks, and is generally more flexible \cite{macleod2022natmat}. 
Despite the increased cost and complexity, there have been several research efforts in this area, spanning applications from autonomous solubility screening~\cite{macleod2020self, Pizzuto2022, Fakhruldeen2022}, photocatalysis~\cite{Burger2020}, and automated synthesis~\cite{Lim2021}.
SDLs provide an exciting semi-structured environment where the robotics community can transfer their methods to novel applications.
Robotics researchers have focused their attention on various applications---from household environments to extreme environments such as nuclear and space---that possess common underlying challenges with SDLs.
For example, the challenges of assistive home robots related to grasping transparent glassware are also present in SDLs.
Towards this goal, learning-based methods e.g. TranspareNet~\cite{xu2021seeing} and MVTrans~\cite{wang2023mvtrans} have been demonstrated for detecting laboratory glassware.
In addition, there exists the need to have more task-specific grippers, such as grippers that are specifically designed for laboratory containers and well-plates~\cite{Zwirnmann2023}.
While  a universal gripper that would exhibit the dexterity of a human hand may seem ideal, we are still quite far from this and  most laboratories have to either adapt their current grippers with 3D printed parts and/or use tool exchangers. 
Alternatively, laboratory tools e.g. pipettes can be made more robot-friendly~\cite{Yoshikawa2023}.
Amongst others, these approaches will pave the way to having more robust laboratory robotics for experiments over long periods of time.

% On when to use robotics - A robotic arm can be overkill (so much has been done with 3d printer technology - i.e., Lilo Pozzo's work).
While there is an increasing interest towards using anthropomorphic robotic platforms, primarily due to their attractive nature of being deployed in human labs, there still exist a large number of challenges with having these platforms carry out long-term experiments.
As a robot can be described as any system of sensors and actuators that can perform a complex task, it is worth noting when a workcell or other automated unit operation can be used to substitute or supplement an anthropomorphic robot.
To this end, modular systems that can---in principle---handle a wider range of experiments due to the closed nature of their subsystems have been demonstrated as alternatives \cite{Steiner2019, Manzano2022}. 
To date, the SOTA in using anthropomorphic robotic systems in SDLs have carried out experiments on open benches, which limits the generalisation of experiments towards materials that have a higher degree of toxicity.  
%While they make snazzy publications, they don't always make more effective science.
%ROS is hard, but great because non-proprietary, capability to get creative (incorporate ML), extensibility.
\par
Democratization of laboratory automation is a crucial path for the contemporary developments of individual lab groups to transcend beyond local, bespoke solutions. 
It has thus far been exceptionally difficult for the community to transfer knowledge between laboratories regarding their hardware development and integration.
We are calling for a focus and investment in modularity and open hardware, that makes use of the aforementioned communication approaches. 
Publishing modular hardware components---either through new journals or maker spaces---will enable the community to take advantage of rapid prototyping and manufacturing.
The combination of open hardware and open control software will have the accelerating effect of democratizing access to lab automation and ensuring our brilliant peers have this technology regardless of resources. 

\subsection{Miniaturization}\label{sec:minaturization}
Notably, we see multiple opportunities in miniaturization and reducing the footprint of hardware to both reduce the barrier to entry and reduce material consumption and waste production. 
Digital microfluid platforms have a high capability for miniaturization because they replace moving mechanical parts with a liquid that needs to be moved \cite{Soldatov2021Self-Driving}. 
The HTE community in biology and chemistry has long used microwell plates to reduce sample volumes to the microliter scale \cite{hertzberg2000high,battersby2002novel}.
In materials science and chemistry,  miniaturization can have a high impact \cite{potyrailo2012high,karthik2016miniaturized} but is to date underexplored.
In addition to adopting microwell plates \cite{buitrago2015nanomole}, microfluidic reactors allow for samples in SDL to reach the nanoliter scale \cite{epps2020artificial}.
SDLs built entirely using flow chemistry can take advantage of established chemical engineering without anthropomorphic robots. 
Flow reactors are also easier to integrate with online characterization techniques \cite{abolhasani2023rise}. 
Milli- or micro-fluidic reactors have the advantages of higher rates of heat, mass, and photon transfers as a result of the enhanced surface-to-volume ratio. 
Nonetheless, flow reactors suffer from the possibility of clogging  when dealing with solid-state materials or precipitates \cite{Delgado-Licona2023Research}.
These cases (including thin-film preparation, battery materials, or polymerization with precipitation of solid products or byproducts) are better suited for parallel batch reactors\cite{abolhasani2023rise}.
Modular microfluidic units could accelerate process optimization and formulation discovery; however, a standardized protocol in modular configuration for a targeted reactive system needs to be established \cite{Volk2022Flow}. 
\par

In addition to the application of microfluidic reactors, miniaturization has also been applied to devices and solids sample arrays. 
This has been approached using combinatorial synthesis in small areas (<1~mm$^2$)\cite{TAKEUCHI2005Combinatorial}, multinary thin film synthesis \cite{ludwig2019discovery, nikolaev2014discovery}, and microdroplet array synthesis \cite{feng2018droplet,rosenfeld2020miniaturized,seifermann2023high}. 
Recent work with scanning probes has shown that sub-femtoliter solutions can be patterned and combined on surfaces, providing further opportunities to miniaturize experiments~\cite{saygin2021closed,brown2022scanning}. 
Such samples can also serve as miniaturized reactors \cite{zech2002miniaturized}, having the broad effects of reducing material consumption and increasing experimental throughput. 
In addition to miniaturizing samples, it is also powerful to minimize the analytical instrument to study these samples, as exemplified by scanning droplet cells to study corrosion \cite{lohrengel2000electrochemical} and adapting electrochemical characterization techniques  \cite{JENEWEIN2022Automated, ZHANG2020Miniaturized}. 
While not all characterization techniques can be applied in miniaturized platforms \cite{abolhasani2023rise}, we encourage further developments in this area that increase automated capacity while reducing material consumption.

%%% Local Variables: 
%%% mode: latex
%%% TeX-master: "./main.tex"
%%% End:
\section{Algorithms}\label{sec:software}

% 1. Whats out there and what's missing
% 2. From optimization to knowledge generation

\subsection{Autonomous decision making for optimization problems}
% here we can write about state-of-the-art, Bayesian optimization based experiments, also with some references to other optimization methods, e.g. GAs etc.

The traditional design of experiments (DoE) becomes rapidly impractical for high dimensional problems due to the exponential growth of the number of required experiments.
Incorporating ML in SDLs has emerged as an efficient way to explore the chemical space and to speed up experimentation.
Taking advantage of the information generated during the optimization process itself, ML enables an iterative experimental design that maximizes the information gained per sample and that requires a smaller number of experiments with respect to traditional DoE \cite{Delgado-Licona2023Research, roch2020chemos}.

At present, a variety of ML approaches have been applied to SDLs.
Genetic algorithms (GA) are a class of adaptive heuristic search algorithms inspired by the process of natural selection that can be used for solving both constrained and unconstrained optimization problems.
GAs have been applied recently to optimize the conditions to produce gold nanoparticles \cite{salley2020nanomaterials}.
Reinforcement learning (RL) has also been successfully applied in SDLs \cite{rajak2021autonomous, li2020autonomous, Maffettone2021Badseed}. This is an ML paradigm that enables an agent to learn through trial and error in an interactive environment by taking actions and receiving rewards, with the goal to learn a policy that maximizes the total reward over time.
The most widely used decision-making algorithm in SDLs is Bayesian optimization (BO), a method particularly suited to balance the trade-off between exploration and exploitation of the input parameter space.
BO has been applied to SDLs both in the single objective setting \cite{macleod2020self, gongora2020bayesian, shields2021bayesian} and, more recently, for the simultaneous optimization of multiple objectives \cite{macleod2022self}. 
There are two main strategies for the implementation of multi-objective BO: combining multiple objectives into one (e.g., Chimera \cite{hase2018chimera}) and identifying a Pareto front that trades off among the multiple objectives (e.g., qNEHVI \cite{daulton2021parallel}).
The second approach has the advantage of not requiring the experimentalist to select the trade-off among the different objectives \emph{a priori}.
\par

A challenge in SDLs is selecting suitable algorithms for each specific scenario. 
Open-source software packages for SDLs offer an easy-to-use starting point for non-experts in machine learning to begin autonomous experimentation (ChemOS \cite{roch2020chemos}, EDBO+ \cite{shields2021bayesian, torres2022multi}). 
Other general-purpose libraries are becoming more user-friendly and are constantly updated with SOTA methods, such as BoTorch \cite{balandat2020botorch}, or \href{https://ax.dev/}{Ax}, an adaptive experimentation platform built on top of BoTorch.
Often, off-the-shelf decision-making algorithms in general require further tuning, which can slow research and even undermine an experimentalist's intent in purpose of applying them.
In this sense, the lack of open-access datasets for experimental campaigns is a current issue.
% Transparency and open resources have been traditionally valued by AI, computational, and machine-learning communities in academic publications.
% The adoption of equivalent community standards in the field of SDLs would be highly beneficial.
Relating back to our discussions around data (\autoref{sec:data_sharing}), the availability of data would allow researchers to evaluate novel algorithms on multiple surrogate systems based on real experiments, while the absence of such datasets could considerably impede the development of dedicated algorithms and create obstacles for the development of autonomous platforms \cite{liang2021benchmarking, epps2021universal}.

An open challenge is the incorporation of generative models \cite{sanchez2018inverse,gomez2018automatic,noh2020machine} in SDLs.
While being very successful in finding hypothetical molecules with tailormade properties, i.e. solving the inverse problem of materials design, generative models frequently suggest molecules with complex or unknown synthesis routes \cite{gao2020synthesizability,bilodeau2022generative}, which limits their real-world impact and prevents their application in automated labs.
Including synthesizability or even synthesis planning in generative models for inverse design, as well as developing versatile multi-objective generative models is a promising path toward their integration in SDLs \cite{gao2022autonomous,NEURIPS2019_46d0671d}.

A further open, yet more technical challenge in self-driving labs is the systematic exploitation of existing data and also process descriptions \cite{kearnes2021open} that are published in the literature.
Automatic extraction of that data and conversion from natural, i.e. informal language to computer-readable, i.e. formal language is an open challenge, currently requiring a large amount of manual work \cite{luo2022mof}.
Large language models can potentially help in that task \cite{gupta2022matscibert,dunn2022structured} but further research is required to reliably extract data and knowledge from scientific literature.

%future directions (Luca Torresi will work on this later, but everyone can add thoughts here!):
%\begin{itemize}
    %\item Up to now, the sheer majority of the algorithms that have been applied for the optimization of SDLs are general purpose, and don’t take advantage of any prior knowledge regarding the process in question. An interesting direction for future research would be the development of methods to embed prior expert/domain knowledge in the decision-making algorithms (e.g., physics informed NN). % this is already covered in the next subsection!
    %\item transfer learning across different platforms (meta-learning?) % this is already covered in the next subsection!
    %\item combine with inverse-design % this is a good idea, we can expand on this more here.
    %\item improve interpretability of ML models % this should also go into the next subsection, as it goes more towards knowledge generation
    %\item making the design space exploration/exploitation software user-friendly % this also fits here, as it is more of a technical point.
%\end{itemize}

\subsection{Knowledge Generation}\label{sec:knowledge_generation}
% How do we enable SDLs to make true scientific discoveries? 
% Not necessarily getting final answers or compiling results but rather getting information that could give other problems a leg up on their efficiency, 

The ultimate goal of scientific experiments is typically not (only) the generation of data, but the generation of knowledge and understanding \cite{krenn2022scientific}.
From that aspect, the main objective of SDLs should go beyond solving optimization problems in high dimensional spaces of materials and processing conditions.
Rather, they should aim to interpret that data, link it with other data (potentially from other SDLs and databases), and help to generate and test scientific hypotheses (potentially in a semi-autonomous or autonomous manner~\cite{friederich2021scientific}). 
To come closer to the goal of autonomous generation of scientific knowledge and understanding,we propose multiple considerations for the future.
\par

Firstly, in order to leverage data generated across labs and use it to train models, further data analysis, and generate new scientific knowledge, we underscore on the efforts highlighted in \autoref{sec:data_sharing} to  link data through shared data formats,  metadata definitions, vocabularies and ontologies. This will move SDLs beyond the discovery of single, interesting data points.
From these innovations, data can be subsequently used for transfer learning models, multi-fidelity and multi-task models, as well as representation learning methods, which can learn from heterogeneous datasets \cite{ProtBERT}.
Such pre-trained models enhance the decision-making process in SDLs.
In particular, they not only learn from locally generated data but already have prior knowledge that enhances decision-making early in an experimental campaign.
%However, if vocabularies, meta-data, and data formats are not globally shared (at least within one domain or community), the exchange cannot be performed automatically but requires manual data collection and adaption.
%This will limit the opportunities to learn from shared knowledge and prevent us from reducing the number of redundant experiments that have to be performed across different labs.
Lastly, we encourage the development of AI/ML methods for SLDs that reach beyond optimization problems. 
SDLs are very good at finding optima, which can act as sources of inspiration for scientific understanding, as defined in Krenn {\it et al.} \cite{krenn2022scientific}. Furthermore, increasingly sophisticated explainable and self-explaining machine learning models for molecules \cite{ying2019gnnexplainer, teufel2022megan}, materials science \cite{oviedo2022interpretable, zhong2022explainable} and particularly SDLs \cite{pilania2021machine, kailkhura2019reliable} pave ways towards autonomous loops of hypothesis generation.
Methods such as automated generation of counterfactuals \cite{wellawatte2022model} offer further opportunities for automated hypothesis testing, when combined with fully automated synthesis and characterization, or accurate predictive simulation workflows.

%%% Local Variables: 
%%% mode: latex
%%% TeX-master: "./main.tex"
%%% End:
\section{Scaling autonomous discovery}\label{sec:scaling}
% Pull exactly from scalability.tex
% initial discussion is focused on motivations for scaling beyond the single-laboratory scale SDL

% opportunity and scope
Taking the opportunities of data, hardware, and software in concert, we can turn our attention to economies of scale. 
In this section, we focus our discussion on scaling via interconnection of multiple distinct hardware modules and SDLs \cite{MaffettoneMASS}, or via increasing the size, capacity, or extent of a given SDL. 
The scalability of SDLs supports a faster, more efficient exploration of experimental parameter space, as well as a larger volume output of manufacturing processes. 
\par

% motivation 1: even higher throughput
SDLs at the laboratory scale already empower HTE increasing throughput by orders of magnitude.
Scaling an SDL beyond the single lab will increase experimental throughput proportionately.
% M2
Moreover, moving from local autonomy to distributed autonomy will enable optimization and search over multiple length scales, characterization methods, and related systems. 
For example, a lab-scale SDL within a single confined system, such as glove box handling of liquids,  may enable a higher experimental throughput. 
Coupling this with an SDL at a different scale, such as a synchrotron, may enable new discoveries by bridging techniques across multiple length scales.
% M3
Scalable computational approaches can capitalize on a greater experimental throughput and the interplay between modular SDLs for a more efficient search through space. 
% M4 
Designing SDLs that act in concert over multiple fidelities and length scales will enable materials verification and validation that is nearer to industrial requirements. 
This ties in with the ``advanced manufacturing'' movement that has received significant attention in the last decade \cite{Cheng_2018}. 
\par

%\subsection{Challenges to scaling autonomy}
% here we call out the R&D challenges that arise specifically from scaling up to multi-SDL, autonomous facilities, distributed SDLs
Building scalable SDLs requires a particular landscape of considerations. Broadly these include the manufacturing considerations and the transition from automation to local autonomy to distributed autonomy.
There are challenges in scaling the volume of experiments, synchronizing data (\autoref{sec:data_sharing}), algorithms for handling multifidelity and multimodal data (\autoref{sec:software}), and software that enables distributed orchestration across platforms.
As the number of experiments performed per unit of time increases, it is crucial that experimental platforms are found that minimize the amount of material required per experiment (\autoref{sec:hardware}).
This reduces the costs of experiments, makes them more amenable to parallelization, and reduces the time required for some types of processing steps.
However, this may change the relevance to manufacturing scale, so there is a case to be made for multi-scale automation that features high-throughput experiments at a highly miniaturized scale and lower-throughput experiments at larger scales.

% challenge 1:
Scaling beyond single-laboratory SDLs creates some open questions in cost analysis.
How do we model the cost of implementing and operating such a large-scale high-throughput discovery architecture?
Can we quantify the aspects of scale that result in cost efficiency?
The capital expenses would include floor space, instrumentation (sample production, characterization, storage), and mechanical infrastructure (robotics, table space, resource garages).
The operational expenses would include materials, power, maintenance, and replacement.
There are broad considerations around architectural design, and efficiently accommodating the disparate instrumentation sets required by different workflows.
We encourage such engineering and feasibility analyses by public research centers to be made broadly available, to inform continual improvement cycles \cite{deming1982quality}.
Calling back to our discussion on modularity, we further highlight the economic analyses of ``platform'' (i.e. incremental)  vs ``bespoke'' (i.e. single-leap) development strategies for large projects, and resoundingly encourage the former \cite{ansar2022solve}.

% challenge 2:
When considering the transition from automation to local autonomy to distributed autonomy, we are focusing first on the distinction between HTE and AE/SDLs, with a second distinction between an isolated SDL and a distributed SDL or network of distributed SDLs \cite{MaffettoneMASS}.
Mobility may be a key component to making the transition from local to distributed autonomy.
As discussed in Section~\ref{sec:hardware}, this can be accomplished by fixed material transport systems (robotic rail systems, custom feeders) or by more generalized mobile robotic technology.

% challenge 3:
Large-scale autonomous experimentation also creates two opportunities for software innovations.
Scaling SDLs will require actional workflow systems---mediated perhaps by powerful workflow languages---designed to enable a description of the requirements for each sub-task in a workflow as well as the specifics needed to connect intermediate products from one sub-task to the input of subsequent sub-tasks.
% challenge 4:
Intelligent search or orchestration that scales is a second opportunity for software innovation in an ongoing area of rich study.
This requires a capacity for evaluating incoming results from experiments carried out in parallel on the work floor, and converting this new data into new experimental queries.
This computational component connects the stream of outputs from in flight experiments back around to the process that injects new experiments into the available resource pool.
It does so in light of all previously collected experimental data, models of the processes being studied, and a growing model of the abstract landscape defined by the problem goal specification.
\par

%%% Local Variables: 
%%% mode: latex
%%% TeX-master: "./main.tex"
%%% End:
\section{Education}
A large span of educational backgrounds is currently required to drive the development of SDLs, ranging from traditional sciences, technology, engineering, and math (STEM) to humanities.
This raises the question of how to approach educating the next generation of students and researchers so that they are prepared to develop and responsibly use automated and self-driving labs.
With this section, we hope to promote discussion amongst educators as to how to integrate the ideas and techniques of SDLs into a curriculum and provide resources that will enable such development. This section is organized into three subsections, (1) the topics that researchers in this field should know, (2) mechanisms for teaching them, and (3) thoughts on how to assess success. On a larger scale, we hope to spur discussion as to what should be considered foundational knowledge in higher level education and how we enable future generations of scientists to contribute to and advance these growing sectors of research and discovery. 

\subsection{Fundamental knowledge underpinning AI-accelerated science} 

As we have already discussed in great detail, SDLs bridge a broad range of topics such as AI, computing, engineering and automation of experiments. Prospective scientists entering the field of acceleration will typically have one domain of expertise but maybe little to no experience in other domains. Thus, it may be challenging for an individual to master all skills necessary in addition to their experience in their application area. The multidisciplinary nature of the field poses challenges to educators in both deciding on the prerequisites and lecturing for an audience with diverse academic backgrounds and research interests. 

The intellectual barrier to entry into AI-accelerated research involves fundamental theoretical knowledge as well as practical skills, both of which must be acquired through training and practice. Specifically, practitioners must understand the fundamental topics of:
\begin{itemize}
\item how and when to conceive of scientific research as an iterative workflow, involving the selection and performance of experiments together with the subsequent analysis of results,
\item how one can go from an existing gap in knowledge to defining critical bottlenecks that limit the speed at which that gap can be filled,
\item the mathematics and computer science  underpinning SDLs, e.g. statistics, probability theory, linear algebra, programming, automated data-analysis, basic machine learning and automated planning/decision-making,
\item lab automation including existing solutions, modular setup of SDLs, communication protocols of automated equipment,
\item and the history of AI- and automation-accelerated research.
\item lab automation including existing solutions, modular setup of SDLs, communication protocols of automated equipment,
\item and the history of AI- and automation-accelerated research.
\end{itemize}

In addition, practitioners should have the following skills:
\begin{itemize}
    \item data management and curation
    \item algorithmic data processing including scripting the extraction, analysis, and presentation of large datasets from possibly heterogeneous sources.
    \item interdisciplinary teamwork combining software, hardware, and domain experts
    \item and fluency across these disciplinary intersections for effective communication with diverse researchers
\end{itemize}

\subsection{Mechanisms for training in AI-accelerated science}

There are a number of approaches to learning and teaching about AI-accelerated science that can appeal to a wide array of educational backgrounds, experience levels, and time commitments.
We envision a multi-faceted approach to training new students and existing researchers in SDL-accelerated research. Here we provide a list of a few such facets that are organized from lowest barrier to entry to those that require more time, expertise, and resource commitment.  

\begin{itemize}
    \item Freely available videos are a great resource for beginners and the community would benefit from repositories of such videos that allow learners to sort and search to find topics of interest among curated or trusted videos. Such methods have been used systematically in engineering education.\cite{Lindsay2021CSUEngineering}
    \item For students excited about active participation, workshops at conferences are an excellent resource, especially those that are paired with large meetings. For instance, the MRS Data Science Tutorial Organizers hosted two machine learning competitions in recent years, one in fall 2021 focusing on active learning and another in fall 2022 on supervised machine learning.\cite{sun2022teaching}
    \item More formal courses can be helpful for some learners and be longer and more substantial than workshops and tutorials. The collaborative nature of the AI-accelerated materials community raises the possibility of jointly developing a course that can have a presence at multiple universities and touch on many material or chemical domains. One example of how such a course may be developed collaboratively by many researchers comes from the area of computational chemistry, where a shared course has been developed between multiple institutions. % The Pan Canadian Graduate Computational Chemistry...     
    \item As a powerful resource for either advanced users of instructors developing curricula, shared digital resources such as datasets and code can be directly shared. While these often require a certain degree of expertise to incorporate, they provide access to powerful techniques. However, it is necessary to efficiently share these with attribution. One avenue for doing this are repositories of materials informatics resources such as REMI: REsource for Materials Informatics\footnote{https://pages.nist.gov/remi/}.
    \item Student internships in companies that develop robotics hardware and software as well as companies that have long-standing experience in advanced automation will help to transfer existing industry know-how into academic environments and apply it to advance SDL technology.
    \item As in many fields, there is no substitute for learning with hands-on experience. SDLs provide some unique opportunities for such hands-on learning. Suitable cost-effective SDLs \citeauthor{Baird2022Minimal} can be adopted by instructors for teaching settings. Alternatively, enthusiastic learners can directly leverage these resources to learn independently. Furthermore, students should be incorporated in existing and currently developed SDLs through thesis projects or research internships, to be exposed to SDL technology as early as possible, and also to transfer knowledge and know-how between labs. 
\end{itemize} 

A number of these examples show the value of openly sharing software, hardware, and data. As such, we view this as a strong encouragement to continue and expand the practice for education in AI-accelerated research.

\subsection{Assessment of Educational activities}

It is important to evaluate the effectiveness of coordinated training and education efforts in the community.
Moreover, the insight from these assessments should be shared to collectively improve practices. 
While there is no single metric that can address all facets of education activities, the following avenues stand out as promising processes to gather and act on feedback.
\begin{itemize}
    \item Workshops and tutorials can be evaluated by quantifying attendance, soliciting feedback, and student outcomes. While attendance itself is not especially important as different venues lend themselves to different scales, the fluctuation of this quantity over time can speak to the impact and reputation of the event. Feedback on educational events should always be solicited following the event when thoughts are still fresh in the minds of the attendees. This can be through informal short surveys administered online. For classes with certificates, filling out feedback can be made a prerequisite for receiving the certificate. Actual learning outcomes are more challenging but more important to measure. The key question is whether the educational efforts led to research action. While open to confirmation bias, a starting point would be to interview successful students to learn what facets of their education were most effective.
    \item Open-source educational resources can be assessed through the degree to which they are accessed. Basic analytics can help determine how many new users are present and how long these users spend with the resources. Such results can help guide the refinement of further resources. That said, there is a difference between useful, popular, and correct, which means that raw user numbers do not tell the whole story. It would be useful to couple such metrics with expert-curated recommendations to highlight effective resources.
    \item One avenue that has already been impactful in machine learning and computer science is the use of competitions. In addition to generating excitement about the field, these can serve as an avenue for assessment in evaluating the results from typical participants. In addition to competitions with a broad scope, in-class competitions have several advantages\cite{Gamarra2021gamification}, including encouraging learning by doing and providing an easy-to-implement platform for evaluating student progress.
\end{itemize}

Just as self-driving labs represent an iterative cycle in which experiments are performed and the outcome is used to learn and choose subsequent experiments, it is important to view pedagogy along the same lines and use assessment to build on and improve past efforts. This reflects the dynamic and evolving nature of this field and our expectation that it will grow and evolve in the years to come.

\section{Ethics and Community}

Research on and with autonomous experiments could have a powerful impact on society.  A development and deployment process that does not include careful planning, broad consultation, competent execution and ongoing adaptation might create long-term harms that outweigh SDLs' benefits. Although anticipating all potential complications is impossible, exploring possible problems---as well as solutions and mitigations---early and frequently could reduce the expected cost of such issues. The space of ethical considerations relevant to SDLs is too broad to canvass comprehensively here, but this section highlights a few key categories.

First, by lowering the cost and increasing the accessibility of scientific R\&D, SDLs could profilerate destructive capabilities as well as research progress. For example, these facilities might enable actors with malicious intent to develop hazardous materials of biological, chemical and nuclear origin. Research has already shown that relatively simple machine learning methods can generate novel toxins that are potentially more fatal than previously-known substances and that do not feature on chemical controls and watch lists, creating new governance challenges \cite{Urbina2022}. SDLs, if not managed carefully, could enable further experimentation along these lines and possibly the large-scale production of dangerous substances  \cite{hickman2022self}.
SDL governance will have to balance responsiveness to these concerns with sensitivity to researcher privacy. Physical and cyber security will also play a critical role, since poorly-secured labs, regardless of the soundness of their governance, could be vulnerable to hijacking by hostile actors.

Second, neglect and AI safety failures could lead to risks similar to those of malicious intent. Equipment malfunctions, insufficient cleaning and maintenance, poor storage practices and so forth might inadvertently create harmful substances, for instance by contaminating a procedure that would otherwise be safe. This issue is of course not unique to SDLs---is a general lab safety concern---but the absence of regular human supervision removes a critical auditing layer. Relatedly, the increased role of automated systems in SDLs raises the importance of addressing AI safety issues: a powerful, unaligned system prone to misinterpreting user requests or unfamiliar with a comprehensive range of lab safety practices, standards and risks could, given access to a well-stocked scientific facility, do tremendous damage by, for instance, mixing volatile substances or developing and dispersing toxins or pathogens \cite{Turchin2020, OBrien2020} These are among the scenarios that most concern AI safety researchers \cite{Koehler2020, Hilton2022}. 

Third, SDLs, like prior automation, could have adverse social and political consequences. Historical parallels from the industrial revolution to the recent rise of the gig economy show that these costs can include unemployment and underemployment, reduced mental health, a sense of diminished community and security, and inequitable economic impacts. These problems in turn can trigger escalatory cycles of political backlash, and can result in regulation that slows technical progress. To minimize the likelihood and impact of such dangers, the SDL community should not only study technical aspects of the technology, but also investigate adjacent social systems and relevant historical precedents. To this end, SDL developers should partner with economists, historians, social activists and stakeholders likely to affected by the development and deployment of these technologies. However, given the impossibility of perfectly predicting complex social systems, individuals working on SDLs should also prioritize ongoing monitoring of and adaptation to unexpected developments.

Fourth, as a form of economic activity that involves industrial components and processes, SDLs could have negative environmental impacts. For instance, a rise in the use of heavy machinery as the costs of experimentation drop might raise carbon emissions, and increasing chemical R\&D might damage local ecosystems. Guarding against these risks will, like addressing social and political consequences, require a combination of foresight (in this case, making the most of the growing environmental science literature) and responsiveness to unexpected developments. Unlike social and political risks, however, environmental issues stand to benefit fairly directly from the research that SDLs could enable \cite{Pyzer-Knapp_2022, Giro2023}. We encourage the SDL community to make climate-related topics a top area of investigation.

Fifth, even if none of the above risks transpire, SDLs could cause harm via incurring inequitable impacts, for instance by concentrating economic gains, additional research prestige, etc. amongst privileged groups. We encourage building working environments that promote equity, support diversity and require inclusion. As a uniquely interdisciplinary community, we celebrate the strengths that derive from differences. In creating working environments where all are welcomed, valued, respected, and invited to participate fully, we will accelerate SDLs' positive impact. The community should incorporate equity and justice in the selection and implementation of education, research, development, policy, and commercialization. This includes openly distributing the results of early-stage research and development in line with FAIR practices, as well as continued commitment to ethical and reproducible research.

Identifying risks does not solve them, but it represents an important first step. Ideally, SDLs could create products and processes that actively counter these dangers, e.g. by enabling people to concentrate on the safest, most enjoyable aspects of discovery, contributing to climate change mitigation and adaptation, and making scientific knowledge and experimentation more equitable and accessible. However, this will not happen without active guidance; realizing this vision will require a concerted effort from the SDL community.

% We resolve to bring these principles and issues to the awareness of our community along with governmental, non-governmental, and private organizations around the world. We support technological ethics education and training for researchers and encourage members of our community to engage in policy development, be that in the form of engagement with elected officials, writing policy briefs and white papers, or trialing self-regulation and other forms of soft law.

\par

\par

%%% Local Variables: 
%%% mode: latex
%%% TeX-master: "./main.tex"
%%% End:
\section{Conclusion}
The field of AE and SDLs has the potential to power a new revolution in the pace and nature of scientific discovery.
As with any revolution, the community shapes the process and outcome.
At this pivotal nascent moment, we acknowledge that there are rich opportunities at every intersection of our community, from software to hardware to education and ethics.
It is crucial that we take deliberate action to ensure that our collective progress as fruitful and positive as possible. 
Important considerations that must be addressed include how we acquire, store, manage, and share our data, as well as how we develop, disseminate, and scale our hardware and software solutions. 
These innovations do not happen in a vacuum and, as such, we have also highlighted the ethical implications of the field and future education and community needs. 
Born out of a diverse discourse and community feedback, we hope this perspective will provide guidelines, encouragement, and facilitate community building.

% Close out

\section{Acknowledgments}
We are grateful to our peers for their gracious and varied contributions to this work. As many ideas and pieces were shared across various media, the organizing authors have been listed first for correspondence, followed by an alphabetical list of all other contributors. 
\par
We would like to also thank Howie Joress for his insightful discussions.
\par

% Add funding acknowledgements. Just list them for now, don't stress on grammar.
P.M. used resources of the National Synchrotron Light Source II, a U.S. Department of Energy (DOE) Office of Science User Facility operated for the DOE Office of Science by Brookhaven National Laboratory under Contract No. DE-SC0012704.,
and resources of a BNL Laboratory Directed Research and Development (LDRD) project 
23-039 "Extensible robotic beamline scientist for self-driving total scattering studies".
P.F. and N.H. acknowledge support by the Federal Ministry of Education and Research (BMBF) under Grant No. 01DM21001B (German-Canadian Materials Acceleration Center).
P.F. and L.T. acknowledge support by the Federal Ministry of Education and Research (BMBF) under Grant No. 01DM21002A (FLAIM).
C.S. acknowledges financial support from VILLUM FONDEN (Grant No. 50405).

\section{Disclaimer}
These opinions, recommendations, findings, and conclusions do not necessarily reflect the views or policies of NIST or the United States Government.

Certain equipment, instruments, software, or materials are identified in this paper for informational purposes.  Such identification is not intended to imply recommendation or endorsement of any product or service by the authors or their respective institutions, nor is it intended to imply that the materials or equipment identified are necessarily the best available for the purpose.

%%% Local Variables: 
%%% mode: latex
%%% TeX-master: "./main.tex"
%%% End:

% \input{data_sharing_incentives.tex}
% \input{scalabiltiy.tex}  
% \input{education.tex}
% \input{ethics.tex}  
% \input{modularity-reusability-interoperability.tex} 
% \input{miniaturization.tex}
% \input{next-generation-robotics.tex}
% \input{competitions.tex} 
% \input{knowledge-generation.tex}

\bibliography{bibliography}

\end{document}